\documentclass[%
 reprint,
amsmath,amssymb,
 aps,prl
]{revtex4-1}

\usepackage{graphicx}
\usepackage{dcolumn}
\usepackage{bm}
\usepackage{amsmath}
\usepackage{comment}

\begin{document}


\title[Submitted Manuscript]{Emitter-Vacuum coupling through a leaky metal nanostructure and the role of dynamics in density of optical states}

\author{Kritika Jain}
\author{Murugesan Venkatapathi}%
 \email{murugesh@iisc.ac.in}
\affiliation{%
 Computational and Statistical Physics Laboratory, Indian Institute of Science, Bangalore, 560012 
}%


\begin{abstract}
We show a break down of the conventional partition of optical states into its radiative and non-radiative parts. Large divergence of experimental observations from current theory in the case of emitters interacting with fully absorbing plasmonic nanoparticles only a few nanometers in dimensions, are now evident. A model of fluctuation-dissipation demands non-local behavior from $limiting$ small metal nanoparticles and proximal metal surfaces. We point that widely used techniques to enhance optical sensing such as surface-enhanced-Raman-spectroscopy (SERS), may not have been viable but for this effect. Qualitatively, this quantum effect seems to present itself only when the classical probability of scattering of an emitted photon by a near-by absorbing nanostructure approaches zero. Hence, though different in origin and scale, this has an interesting analogy with quantum effects resulting in Hawking radiation near a black-hole.\\
%
\end{abstract}

\pacs{Valid PACS appear here}
\maketitle


\section{Introduction}\label{intro}

Spontaneous emission is typically elucidated as a function of vacuum fluctuations and proximal matter, and its nature in the weak vacuum-coupling regime is predicted by the $density$ of optical states. Note that the strong vacuum-coupling regime becomes relevant when only a few optical states (modes) of vacuum are available for the emission, as in a cavity. In most other applications, increasing density of radiative states relative to the non-radiative states using appropriate nanostructures placed near emitters can result in significant gains in power and even efficiency of emission; and this is broadly referred to as Purcell enhancement \cite{purcell1946purcell,drexhage1970influence,kuhn2006enhancement,bharadwaj2007spectral,cheng2007separation,kim2013precise,dulkeith2002fluorescence,dulkeith2005gold}. This work shows a break down of the current theory of local density of optical states when the dynamics of fluctuations between the emitter and a nanostructure play a significant role in the emission process. Especially, the significant gains in emission observed in presence of resonant and off-resonant fully absorbing metal nanoparticles of limiting small dimensions ($<$ 15 nm)\cite{tripathi2013plasmonic,kang2011fluorescence,schneider2006distance,haridas2013photoluminescence,haridas2011photoluminescence,haridas2010photoluminescence,haridas2010controlled}, directly contradicts our current understanding. A few theoretical evaluations were available in their original reports but this divergence was probably overlooked, while our evaluations are presented here. In conjunction, we also report predictions of our proposed theory that remove any contradictions and divergence.

We also show that widely used techniques such as surface-enhanced-Raman-scattering (SERS) exhibit gains up to $10^{10}$ in magnitude \cite{kovacs1986distance,ye1997surface,mcfarland2005wavelength,dieringer2006introductory,wang2013wafer,gabudean2012gold,ivanov2016strong,kennedy1999determination} due to this effect discussed in this work, which otherwise would be restricted to a factor of $10^3$ at most. In many cases of sensing fluorescence and Raman signals, near-field enhancement of incident radiation exciting the emitter, accompanies a possible enhancement of its emission. Typically a metallic surface or a larger nanostructure hosts smaller (sharper) nano features that result in this near-field enhancement of exciting radiation, but by conventional theory, this surface adds non-radiative states significantly more than radiative states for emission. The observed factors of enhancements due to a metallic surface are much greater than the combined amplification possible due to relative increase in exciting radiation and density of radiative states, for both resonant and off-resonant emissions. The large enhancements of Raman signals observed has so far lead to tentative mechanisms proposed that do not submit to classical electrodynamics and quantum models \cite{le2006rigorous}. This divergence of experiments from theory was mostly attributed to a $chemical$ enhancement of unknown origin \cite{fromm2006SERS}. But later studies with varying chemical properties of dyes showed that these introduce variations of at most an order of magnitude in the surface enhanced Raman signals \cite{sharma2012SERSreview,kneipp2016SERSrigorous}. Further, measurements resolved with a varying distance from the nanostructure \cite{kovacs1986distance,ye1997surface} have shown that the SERS effect has a longer range than that is possible by any chemical effect.
 
In both strong vacuum-coupling \cite{raimond1982statistics,zhu1990vacuum,boca2004observation,yoshie2004vacuum,bernardot1992vacuum} and recently explored strong matter-coupling \cite{bellessa2004matter, zengin2015matter} regimes of emission, experiments establish the splitting of the emitted energy spectrum due to Rabi oscillations. This reversible exchange of the excitation between two oscillators, and appearance of the two possible modes of the coupled system, is observable when decay rates of the two oscillators are relatively small. But a related and significant unknown has been the degree of absorption possible in matter strongly coupled to the emitter. We show that when probability of dissipation of the excitation in the metal nanostructure is significantly larger than the probability of its exchange with the emitter, conventional partition of optical states holds. Else, an inhibited dissipation results, with an equivalent increase in stimulation of the emitter by virtual photons. This behavior can also be modeled as fluctuations in the metal nanostructure leaking to the emitter due to the relatively high rate of exchange of virtual excitations i.e a leaky metal particle. It is predicted for limiting small dimensions of a metal nanoparticle, and a metal structure separated by a few nanometers from the emitter. Thus this regime is well distinguished from Purcell enhancement where the emitter is at resonance with a relatively large plasmonic particle 50 - 200 nm in dimensions, and is separated by distances on the order of its dimensions or larger. There, the interaction is dominated by elastic scattering, adding significantly to the relative density of radiative states as in the conventional theory.

From a classical electrodynamics point of view, immediate near-field zones of irradiated metallic structures and the dissipation by extremely small metal nanoparticles are dominated by evanescent fields. Originally thought to have no physical significance, such fields were found to play a dominant role in phenomena ranging from tunneling across a thin potential barrier \cite{adam93evanescent, meixner94evanescent, baena05evanescent}, to the large fluctuation-driven heat transfer very near a surface \cite{loomis94fluctuation, rytov53fluctuation}. In this alternate elucidation, an absorbing nanostructure coupled to the proximal emitter by its evanescent fields allows tunneling through of photons from the emitter into vacuum. This is an anomalous case of enhanced proximal emission, where conventional theory predicts large non-radiative absorption and a near-zero probability of classical scattering of the emitted photon by the nanostructure. Hence the described effect and its observations have an interesting analogy with the quantum tunneling resulting in Hawking radiation to reveal an otherwise completely absorbing black-hole \cite{Hawking_radiation}.

\section{Results and Discussion}
\begin{figure}
        \begin{center}
            \includegraphics [scale=0.60]{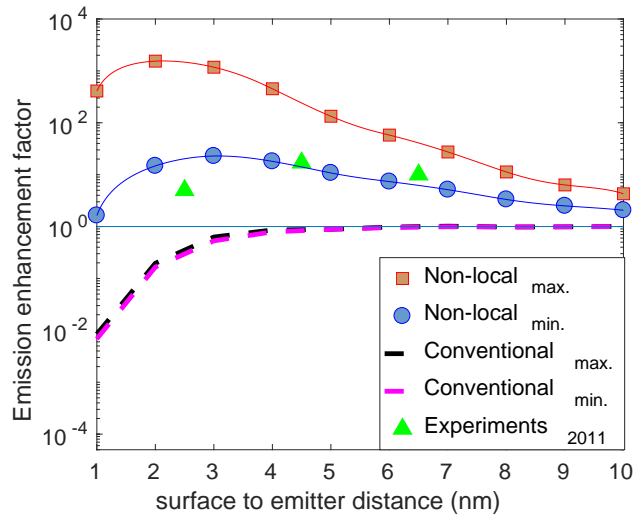}

            \caption{Comparison of theoretical and experimental results: The minimum and maximum gains predicted are given by quantum efficiency $Q$ and power radiated $Q\Gamma^r$, normalized by the values of isolated emitter; note $log$-scale in Y-axis. Experimental data\cite{kang2011fluorescence} is for a gold nanoparticle of radius 5 nm coated with a bi-polymer and the peak emission wavelength was 830 nm.}  
        \end{center}
    \end{figure}

An increase in radiative states was inferred as the increase in vacuum fluctuations coupled to the emitter due to the proximal nanostructure. The smaller nanoparticles can absorb light in the plasmonic range increasing non-radiative states notably, but have a negligible scattering efficiency and do not add to the density of radiative states. But they have been observed to increase rates of spontaneous emission from proximal emitters significantly more than theoretical evaluations, and in cases even notably increase their efficiency of emission in a direct contradiction with theory. Moreover, we point that the same applies in a lesser degree to an emitter separated from surface of a large plasmonic structure by a few nanometers. This effect thus becomes crucial for techniques such as surface-enhanced-Raman-spectroscopy (SERS), which would not be otherwise viable.

     \begin{figure}
        \begin{center}
            \includegraphics [scale=0.60]{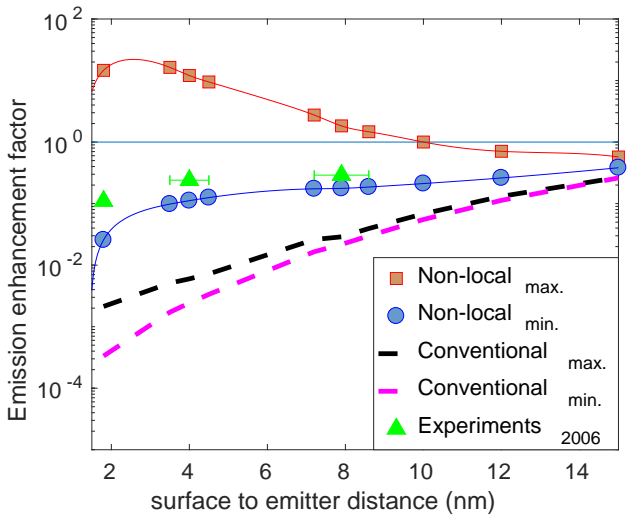}
            \caption{Comparison of theoretical and experimental results: The minimum and maximum gains predicted are given by quantum efficiency $Q$ and power radiated $Q\Gamma^r$, normalized by the values of isolated emitter; note $log$-scale in Y-axis. Experimental data\cite{schneider2006distance} is for gold nanoparticles of radius 6.5 nm and the peak emission wavelength was 520 nm.}
        \end{center}
    \end{figure} 

  \begin{figure}
        \begin{center}
            \includegraphics [scale=0.60]{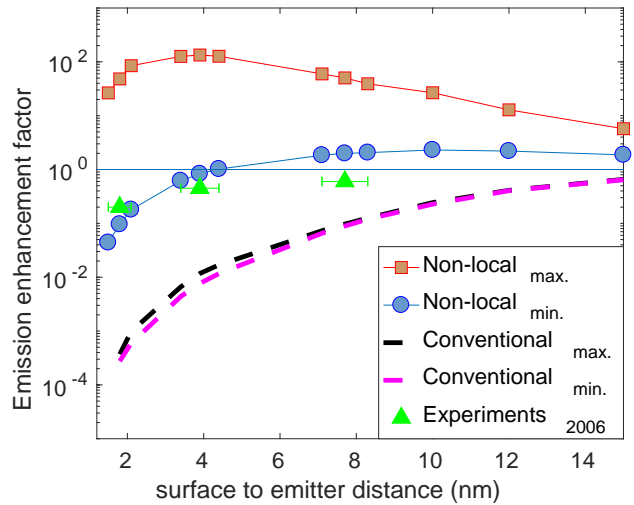}
             \caption{Comparison of theoretical and experimental results: The minimum and maximum gains predicted are given by quantum efficiency $Q$ and power radiated $Q\Gamma^r$, normalized by the values of isolated emitter; note $log$-scale in Y-axis. Experimental data\cite{schneider2006distance} is for gold nanoparticles of radius 6.5 nm and the peak emission wavelength of was 580 nm.}
        \end{center}
    \end{figure} 

First, we begin with some of the experiments of fluorescence and photo-luminescence in materials with metal nanoparticles of limiting small sizes, referred to in the introduction earlier. As this work is not concerned with the type of emitters and preparation of materials, such details of the original experiments are summarized in the supplementary for convenience of the more interested reader. The exciting radiation was off-resonant, and the limiting small absorbing nanoparticle does not significantly alter intensity of light exciting the emitter in these cases. The rates of radiative and non-radiative processes were estimated using independent life-time measurements in many cases, in addition to the measurement of gain in photons emitted. The measured gains are relatively robust and repeatable, while the decay rate is much more sensitive to any uncertainity in distances\cite{Supplementary}. To compare, we present evaluations of both relative quantum efficiency $Q$ and relative power of emission $Q\Gamma^r$, with respect to isolated emitters not interacting with metal nanostructures. The former represents the increase in probability of radiative decay of the excited emitter, while the latter also includes relative increase in ground-state population for emitters that can be excited continuously. Figure 1 presents the measured gains for varying separations of a low efficiency emitter ($Q_o$ $\approx$ 0.012) from a gold nanoparticle of 5 nm radius. The large gains in the observed emission and its contradiction with conventional theory that predicts quenching, are clear. Whereas the proposed $non$-$local$ theory predicts these measurements reasonably well notwithstanding possible uncertainty in efficiency of the isolated emitters. 

Figure 2 and Figure 3 present experiments of quenched emission from two different types of emitters. Here the nanoparticles are marginally larger at 6.5 nm radii and the quenching in the experimental results is significantly less than the expected values of the conventional theory. In contrast these experiments have a reasonable agreement with the predicted range of this extended theory. Note that quenching due to polymer molecules binding the emitter-nanoparticle system has not been included in both the theoretical models of relative emissions. When this is included, both theoretical predictions should reduce notably at larger distances due to the larger effect of polymer molecules. Thus one can infer a better agreement of experiments with the proposed theory, than reflected by figure 3.

     \begin{figure}
        \begin{center}
            \includegraphics [scale=0.60]{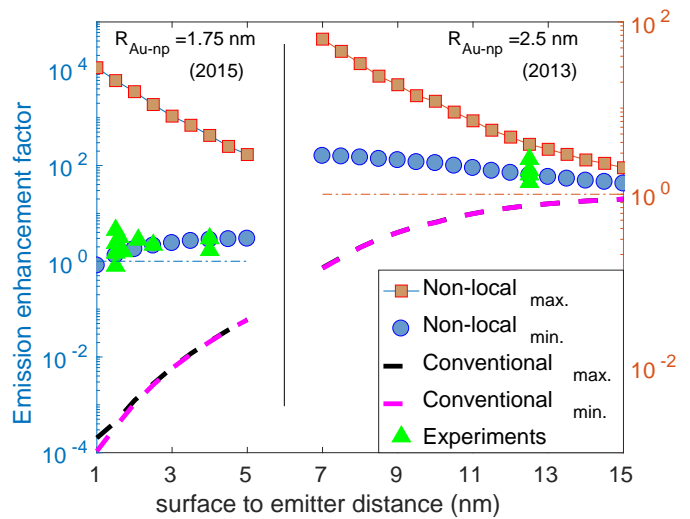}
              \caption{Comparison of theoretical and experimental results: The minimum and maximum gains predicted are given by quantum efficiency $Q$ and power radiated $Q\Gamma^r$, normalized by the values of isolated emitter; note $log$-scale in Y-axis. For short distances 1-5 nm, experimental data \cite{praveena2015plasmon} is for gold nanoparticles of radius 1.75 nm and the peak emission wavelength is 560 nm. For larger distances (7-15 nm) on the right side of the figure, four experimental data points at a mean distance of 12.5 nm \cite{haridas2013photoluminescence} are for gold nanoparticles of radius 2.5 nm and a peak emission wavelength of 560 nm.}
        \end{center}
    \end{figure}

Multiple experiments by other researchers on even smaller gold nanoparticles using self-assembled films and monolayers with quantum dot emitters, are summarized in Figure 4. Note that for the smaller metal nanoparticles discussed in Figures 1 and 4, the experiments present a direct contradiction; theory predicts quenching while experiments report significant enhancements of emission. Monolayers with smaller separations of emitters and gold nanoparticles of a smaller radii of 1.75 nm, have larger divergence with conventional theory as expected (section on the left in figure 4). But the notable enhancement up to a factor of 3 for the dots embedded in films due to sparsely doped and well separated 2.5 nm radii gold nanoparticles are equally unexpected (on the right in figure 4). In this section on the right, four experiments with nearly overlapping gain values are marked along with the predictions of theory. Our modified theory of local density of optical states predicts these experiments reasonably well. One notable aspect of the experiments with limiting small metal nanoparticles of 1.75 nm radii, is the weak sensitivity to number-ratio of emitters and metal particles in the monolayer, and this is not discussed here. The significant effect of virtual plasmons in this case may necessitate the inclusion of any collective behavior among emitters, that has been suggested in the presence of plasmonic nanoparticles \cite{pustovit2009cooperative}, phonon interactions \cite{potma1998exciton} and otherwise at low temperatures \cite{gross1982superradiance}. But these theoretical works involved long-wavelength approximations and more importantly neglected any thermal effects; while other experimental indications of such collective behavior of emitters in the weak vacuum-coupling regime are few so far \cite{scheibner2007superradiance, Goban2015superradiance, Juan2017cooperative}. This weaker additional effect will be addressed elsewhere.
  
      \begin{figure}
        \begin{center}
            \includegraphics [scale=0.60]{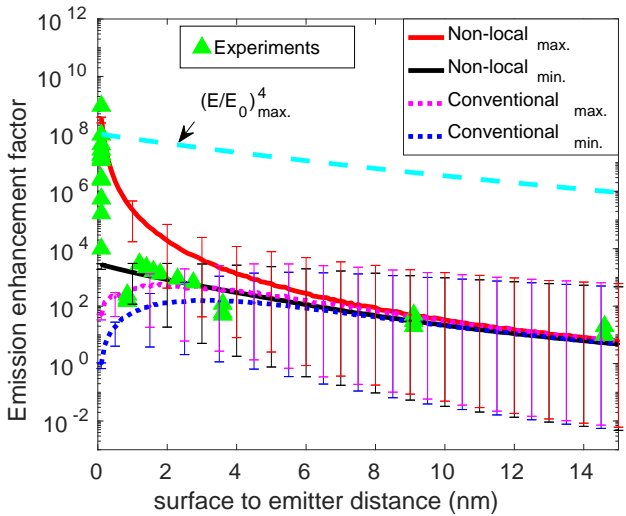}
             \caption{Comparison of theoretical and experimental results: The minimum and maximum gains predicted are given by effective values of quantum efficiency $Q\frac{E^2}{E_o^2}$ and power radiated $Q\Gamma^r\frac{E^2}{E_o^2}$, normalized by the values of isolated emitter; note $log$-scale in Y-axis. Relative $\epsilon = -1.05+0.035i$ for the half-space forming the surface. Error bars placed alternately for conventional and non-local theories reflect variations in predictions using near-field enhancements by 2 to 50 nm size features on the metallic surface, where the lines represent a 10 nm feature.} 
        \end{center}
    \end{figure} 
     
In the case of SERS, we summarize multiple observations to compare predictions of the proposed theory. These enhancements due to metallic nanostructures of various geometries were typically predicted by the power-4 rule i.e. $(E_{local}/E_o)^4$ where subscripts `$o$' and `$local$' differentiate the amplitude of electric field of incident plane wave used for exciting the emitters, and its enhanced value at the location of emitter. Note that difference between excitation and emission energies are not large in SERS measurements, and we ignore this difference in the following interpretation of this rule. The enhancement of the local intensity near a metal structure is given by $(E_{local}/E_o)^2$, and it indicates the increase in probability of excitation of emitters. But for the power-4 rule to be valid in general, we have to infer an additional enhancement of emission by the same factor $(E_{local}/E_o)^2$. Interestingly, the immediate near-field of an absorbing body includes evanescent fields, and this factor represents the cumulative increase in radiative and non-radiative states for such proximal emitters. Thus SERS experiments can be inferred to demonstrate a significant loss of distinction between the radiative and non-radiative states for proximal emitters that are predominantly excited; as predicted by the proposed theory described in the next section. Figure 5 shows the predictions and experimental measurements of a few distance dependent SERS measurements \cite{kovacs1986distance,ye1997surface}, along with a few reported experiments without any spacers between the SERS surface and the emitters\cite{wang2013wafer,gabudean2012gold,ivanov2016strong,mcfarland2005wavelength}. The latter are marked closest to a distance of zero where the probability of excitation is highest. We have plotted theoretical predictions of emission near a plane surface at plasmon resonance (relative permittivity $\epsilon \approx -1$) with near-field enhancement of the excitation due to nano features ranging in size from 2 to 50 nm marked by error bars. The effect of size of these nano features on the fluctuations ($\Gamma_{leak}$ in the next section) were ignored. These represent realistic gold or silver surfaces used in SERS. All experiments are within the predicted range of the non-local theory, but note that the very large Raman enhancements observed are prohibited in the conventional theory. A closer view of the short-range results, off-resonant predictions to account for some of these measurements being off-resonant, and a comparison of the quasi-static and full-wave solutions, are relegated to the supplementary\cite{Supplementary}.

Overall, the proposed theory removes the large divergence observed in the measured enhancements of Raman signals over many decades, and is also supported by the more recent experiments with metal nanoparticles of limiting small dimensions. While our detailed evaluations showing this effect involve dynamics of the strong coupling regime, we believe our alternate elucidation that involves tunneling of photons from emitters coupled to evanescent fields, is also relevant. Further sensitive experiments may firmly establish the full theory defining density of optical states, especially in the strong matter-coupling regime.

\section{Methods: A non-local theory for density of optical states}
Before introducing the proposed extension to theory of density of optical states, we first briefly refer to known methods used to evaluate the modified self-energy of a point dipole in an inhomogeneous medium, and relegate the full details to the supplementary material\cite{Supplementary}. This problem may not have closed-form analytical solutions in general, and a quasi-static solution in the long wavelength limit has been typically preferred \cite{pustovit2010plasmon}. As noted before, this may result in errors in the estimated energy shifts \cite{van2012spontaneous}, and in a underestimation of non-radiative decay rates in metal nanostructures \cite{venkatapathi2014collective}. Hence, we have resorted to more computationally intensive approaches to include retardation for finite wavelengths, and evaluate the variables in the proposed theory described below. But we emphasize that the divergence of experiments from conventional theory is evident even in calculations using first approximations. Single dipole approximation of these smaller metal nanoparticles, or an quasi-static approximation of a metal surface in the case of SERS experiments, are also presented in the supplementary \cite{Supplementary}.

The additional decay rate due to the metallic structure is given by the imaginary part of its self-energy contribution $\Sigma$ i.e. $\Gamma=-2\Im(\Sigma)$; in units normalized by the reduced Planck's constant as in equation \eqref{energy_shift}. The contribution of the nanostructure to self-energy of an emitter at $\mathbf{r_o}$ is given by:

\begin{equation} \label{self_energy}
\Sigma(\omega) = \frac{-2\pi q^2\omega}{mc^2}\mathbf{e}_1 \cdot \mathbf{G} (\mathbf{r_o,r_o};\omega) \cdot \mathbf{e}_1
\end{equation}

and the above can be integrated over $\mathbf{e}_1$ for an average over all polarization, and over frequency $\omega$ with the relative spectral density of the free-space emitter in case of broad-band emission. Here $q$ is the oscillating charge, $m$ is its mass, and $c$ is speed of light. The evaluation of the required dyadic Green tensors $\mathbf{G}$ for an arbitrary structure is explained in the supplementary. $\Gamma^r_0$ and $\Gamma^{nr}_0$ are known radiative and non-radiative decay rates of the isolated emitter adding to $\Gamma_o$. The total radiative and non-radiative parts are a sum of the free-space and metallic components as below.

\begin{equation}
\Gamma^r_{total}=\Gamma^r_0+\Gamma^r 
\end{equation}

Note that a dipole oscillator with an energy of one quantum represents the emitter as a two-level system in this weak vacuum-coupling regime. Then it is convenient to drop charge, mass and amplitude of the oscillator and normalize all self-energy components by $\Gamma^r_0$ for evaluations, where $\Gamma^r_0=\frac{2\sqrt{\epsilon_o}\mu^2k^3}{3\hbar}$ and $\mu$ is the electric dipole moment of the emitter; $k$, $\epsilon$ and $\hbar$ are the wave number, free-space permittivity and reduced Planck's constant.

\begin{equation}
\Gamma^{nr}_{total}=\Gamma^{nr}_0+\Gamma^{nr} 
\end{equation}

where  $\Gamma^r$ and $\Gamma^{nr}$ are additional radiative and non-radiative decay rates of emitter in the presence of metal nanostructure, adding to the total metallic contribution $\Gamma$. The optical theorem for a point source establishes that the evaluated self-interaction of an emitter due to a proximal body represents the total radiative (scattering) and non-radiative (absorption) states of the body. The spatial reflection-symmetry of the free-space and the scattering component of the Green dyads, allows us to equally interpret this perturbation to self-energy as additional action of vacuum on the emitter due to presence of the body. In the rotating wave approximation ($|\Re(\Sigma)| << \omega$), the real part of self-energy in equation \eqref{self_energy} represents the energy shift in the emission due to the nanostructure \cite{pustovit2010plasmon}.

\begin{equation} \label{energy_shift}
    \hbar \Re(\Sigma)=\Delta E
\end{equation}

More importantly, $|\Delta E|/\hbar$ is the rate of exchange of excitation between emitter and nanostructure i.e. the frequency of Rabi oscillations. This exchange in the form of emission and absorption of virtual photons is accompanied by the creation and annihilation of virtual plasmons in the metal particle. Typically, $\Gamma >> |\Delta E|/\hbar$ resulting in a negligible probability of a virtual plasmon in metals, and this regime is also irrelevant for dielectric materials where $\Gamma^{nr}$ is negligible. But in case of the smallest metal nanoparticles, decay rates and frequency of Rabi oscillations are comparable. Also, for an emitter very close to a large metallic nanostructure the above inequality weakens significantly. Note that this would replace absorption with an equivalent generation of photons by virtual plasmon annihilation, and is especially significant for the dipole mode of a nanostructure that represents most of its coupling to vacuum modes. These fluctuations of the dipole mode can not be distinguished from its scattering of virtual photons from vacuum. Only the latter contribution of elastic scattering is included in the increase of radiative states in conventional theory. Thus, the following extension of the local density of optical states becomes necessary.

First, we assume the memory-less probability $density$ function in time $\tau$ for the real decay of plasmon i.e. $\Gamma e^{-\Gamma \tau}$.  A typical case of integration of this probability density in the interval [0,t], results in a decay of an excited state as $e^{-\Gamma t}$ for all $t$. But here we restrict it to the interval of a Rabi oscillation as shown below.

\begin{equation} \label{eq:large delta}
     \mathcal{P}_{real} = \int_{\tau=0}^{\tau=\frac{\hbar}{|\Delta E|}} \Gamma e^{-\Gamma \tau} d\tau
\end{equation}

Since the virtual decay of plasmon is due every Rabi oscillation between the emitter and the nanoparticle, we use the complement of $\mathcal{P}_{real}$ as the probability of generation of virtual photons. This probability of creation and annihilation of virtual plasmons can be non-negligible and thus given by:

 \begin{equation}
      \mathcal{P}_{virtual} = 1 - \mathcal{P}_{real} = e^{-\frac{\hbar \Gamma}{|\Delta E|}}
 \end{equation}

The dipole mode contributions are numbered '1' in the subsequent part of this section. The marginally weaker absorption of higher order modes of an arbitrary nanostructure, made possible due to this dynamics is not discussed, considering we are interested only in limiting small nano-spheres and image dipoles due to surfaces. The effect of additional fluctuations of the dipole mode of the nanostructure on the rates of emission is trivially evaluated using work done on the emitter, as in the classical optical theorem:
 
\begin{equation}
    \Gamma_{leak}=e^{-\frac{\hbar \Gamma}{|\Delta E|}}\cdot\Gamma^{nr}_1
\end{equation}

This rate accounts to a vanishing of non-radiative absorption of the dipole mode and its appearance as a stronger radiative mode. The effective decay rates thus become:

\begin{equation}
\Gamma^r_{eff}=\Gamma^r_0+\Gamma^r+\Gamma_{leak}
\end{equation}
and
\begin{equation}
\Gamma^{nr}_{eff}=\Gamma^{nr}_0+\Gamma^{nr}-\Gamma_{leak}
\end{equation}

The observed quantum efficiency is then:
\begin{equation}
    Q=\frac{\Gamma^r_{eff}}{\Gamma^r_{eff}+\Gamma^{nr}_{eff}}
\end{equation}

A salient point in the above corrections is the strong effect even a relatively small $\mathcal{P}_{virtual}$ has on the radiative rates and the quantum efficiency, in case of a proximal large metallic nanostructure or surface dominated by non-radiative decay (i.e. $\Gamma^r_{eff}$ $\gg$ $\Gamma^r$+$\Gamma^r_0$).

 \bibliographystyle{apsrev}

\begin{thebibliography}{36}
\expandafter\ifx\csname natexlab\endcsname\relax\def\natexlab#1{#1}\fi
\expandafter\ifx\csname bibnamefont\endcsname\relax
  \def\bibnamefont#1{#1}\fi
\expandafter\ifx\csname bibfnamefont\endcsname\relax
  \def\bibfnamefont#1{#1}\fi
\expandafter\ifx\csname citenamefont\endcsname\relax
  \def\citenamefont#1{#1}\fi
\expandafter\ifx\csname url\endcsname\relax
  \def\url#1{\texttt{#1}}\fi
\expandafter\ifx\csname urlprefix\endcsname\relax\def\urlprefix{URL }\fi
\providecommand{\bibinfo}[2]{#2}
\providecommand{\eprint}[2][]{\url{#2}}

\bibitem[{\citenamefont{Purcell}(1946)}]{purcell1946purcell}
\bibinfo{author}{\bibfnamefont{E.~M.} \bibnamefont{Purcell}},
  \bibinfo{journal}{Phys. Rev.} \textbf{\bibinfo{volume}{69}},
  \bibinfo{pages}{681} (\bibinfo{year}{1946}).

\bibitem[{\citenamefont{Drexhage}(1970)}]{drexhage1970influence}
\bibinfo{author}{\bibfnamefont{K.}~\bibnamefont{Drexhage}},
  \bibinfo{journal}{J. Luminescence} \textbf{\bibinfo{volume}{1}},
  \bibinfo{pages}{693} (\bibinfo{year}{1970}).

\bibitem[{\citenamefont{K{\"u}hn et~al.}(2006)\citenamefont{K{\"u}hn,
  H{\aa}kanson, Rogobete, and Sandoghdar}}]{kuhn2006enhancement}
\bibinfo{author}{\bibfnamefont{S.}~\bibnamefont{K{\"u}hn}},
  \bibinfo{author}{\bibfnamefont{U.}~\bibnamefont{H{\aa}kanson}},
  \bibinfo{author}{\bibfnamefont{L.}~\bibnamefont{Rogobete}}, \bibnamefont{and}
  \bibinfo{author}{\bibfnamefont{V.}~\bibnamefont{Sandoghdar}},
  \bibinfo{journal}{Phys. Rev. Lett.} \textbf{\bibinfo{volume}{97}},
  \bibinfo{pages}{017402} (\bibinfo{year}{2006}).

\bibitem[{\citenamefont{Bharadwaj and Novotny}(2007)}]{bharadwaj2007spectral}
\bibinfo{author}{\bibfnamefont{P.}~\bibnamefont{Bharadwaj}} \bibnamefont{and}
  \bibinfo{author}{\bibfnamefont{L.}~\bibnamefont{Novotny}},
  \bibinfo{journal}{Opt. Exp.} \textbf{\bibinfo{volume}{15}},
  \bibinfo{pages}{14266} (\bibinfo{year}{2007}).

\bibitem[{\citenamefont{Cheng and Xu}(2007)}]{cheng2007separation}
\bibinfo{author}{\bibfnamefont{D.}~\bibnamefont{Cheng}} \bibnamefont{and}
  \bibinfo{author}{\bibfnamefont{Q.-H.} \bibnamefont{Xu}},
  \bibinfo{journal}{Chem. Commun.} \textbf{\bibinfo{volume}{0}},
  \bibinfo{pages}{248} (\bibinfo{year}{2007}).

\bibitem[{\citenamefont{Kim et~al.}(2013)\citenamefont{Kim, Yokota, Taniguchi,
  and Nakayama}}]{kim2013precise}
\bibinfo{author}{\bibfnamefont{D.}~\bibnamefont{Kim}},
  \bibinfo{author}{\bibfnamefont{H.}~\bibnamefont{Yokota}},
  \bibinfo{author}{\bibfnamefont{T.}~\bibnamefont{Taniguchi}},
  \bibnamefont{and} \bibinfo{author}{\bibfnamefont{M.}~\bibnamefont{Nakayama}},
  \bibinfo{journal}{J. Appl. Phys.} \textbf{\bibinfo{volume}{114}},
  \bibinfo{pages}{154307} (\bibinfo{year}{2013}).

\bibitem[{\citenamefont{Dulkeith et~al.}(2002)\citenamefont{Dulkeith, Morteani,
  Niedereichholz, Klar, Feldmann, Levi, Van~Veggel, Reinhoudt, M{\"o}ller, and
  Gittins}}]{dulkeith2002fluorescence}
\bibinfo{author}{\bibfnamefont{E.}~\bibnamefont{Dulkeith}},
  \bibinfo{author}{\bibfnamefont{A.}~\bibnamefont{Morteani}},
  \bibinfo{author}{\bibfnamefont{T.}~\bibnamefont{Niedereichholz}},
  \bibinfo{author}{\bibfnamefont{T.}~\bibnamefont{Klar}},
  \bibinfo{author}{\bibfnamefont{J.}~\bibnamefont{Feldmann}},
  \bibinfo{author}{\bibfnamefont{S.}~\bibnamefont{Levi}},
  \bibinfo{author}{\bibfnamefont{F.}~\bibnamefont{Van~Veggel}},
  \bibinfo{author}{\bibfnamefont{D.}~\bibnamefont{Reinhoudt}},
  \bibinfo{author}{\bibfnamefont{M.}~\bibnamefont{M{\"o}ller}},
  \bibnamefont{and} \bibinfo{author}{\bibfnamefont{D.}~\bibnamefont{Gittins}},
  \bibinfo{journal}{Phys. Rev. Lett.} \textbf{\bibinfo{volume}{89}},
  \bibinfo{pages}{203002} (\bibinfo{year}{2002}).

\bibitem[{\citenamefont{Dulkeith et~al.}(2005)\citenamefont{Dulkeith, Ringler,
  Klar, Feldmann, Munoz~Javier, and Parak}}]{dulkeith2005gold}
\bibinfo{author}{\bibfnamefont{E.}~\bibnamefont{Dulkeith}},
  \bibinfo{author}{\bibfnamefont{M.}~\bibnamefont{Ringler}},
  \bibinfo{author}{\bibfnamefont{T.}~\bibnamefont{Klar}},
  \bibinfo{author}{\bibfnamefont{J.}~\bibnamefont{Feldmann}},
  \bibinfo{author}{\bibfnamefont{A.}~\bibnamefont{Munoz~Javier}},
  \bibnamefont{and} \bibinfo{author}{\bibfnamefont{W.}~\bibnamefont{Parak}},
  \bibinfo{journal}{Nano Lett.} \textbf{\bibinfo{volume}{5}},
  \bibinfo{pages}{585} (\bibinfo{year}{2005}).

\bibitem[{\citenamefont{Tripathi et~al.}(2013)\citenamefont{Tripathi, Praveena,
  and Basu}}]{tripathi2013plasmonic}
\bibinfo{author}{\bibfnamefont{L.}~\bibnamefont{Tripathi}},
  \bibinfo{author}{\bibfnamefont{M.}~\bibnamefont{Praveena}}, \bibnamefont{and}
  \bibinfo{author}{\bibfnamefont{J.}~\bibnamefont{Basu}},
  \bibinfo{journal}{Plasmonics} \textbf{\bibinfo{volume}{8}},
  \bibinfo{pages}{657} (\bibinfo{year}{2013}).

\bibitem[{\citenamefont{Kang et~al.}(2011)\citenamefont{Kang, Wang, Jasinski,
  and Achilefu}}]{kang2011fluorescence}
\bibinfo{author}{\bibfnamefont{K.~A.} \bibnamefont{Kang}},
  \bibinfo{author}{\bibfnamefont{J.}~\bibnamefont{Wang}},
  \bibinfo{author}{\bibfnamefont{J.~B.} \bibnamefont{Jasinski}},
  \bibnamefont{and} \bibinfo{author}{\bibfnamefont{S.}~\bibnamefont{Achilefu}},
  \bibinfo{journal}{J. Nanobiotechnol.} \textbf{\bibinfo{volume}{9}},
  \bibinfo{pages}{16} (\bibinfo{year}{2011}).

\bibitem[{\citenamefont{Schneider et~al.}(2006)\citenamefont{Schneider, Decher,
  Nerambourg, Praho, Werts, and Blanchard-Desce}}]{schneider2006distance}
\bibinfo{author}{\bibfnamefont{G.}~\bibnamefont{Schneider}},
  \bibinfo{author}{\bibfnamefont{G.}~\bibnamefont{Decher}},
  \bibinfo{author}{\bibfnamefont{N.}~\bibnamefont{Nerambourg}},
  \bibinfo{author}{\bibfnamefont{R.}~\bibnamefont{Praho}},
  \bibinfo{author}{\bibfnamefont{M.~H.} \bibnamefont{Werts}}, \bibnamefont{and}
  \bibinfo{author}{\bibfnamefont{M.}~\bibnamefont{Blanchard-Desce}},
  \bibinfo{journal}{Nano Lett.} \textbf{\bibinfo{volume}{6}},
  \bibinfo{pages}{530} (\bibinfo{year}{2006}).

\bibitem[{\citenamefont{Haridas et~al.}(2013)\citenamefont{Haridas, Basu,
  Tiwari, and Venkatapathi}}]{haridas2013photoluminescence}
\bibinfo{author}{\bibfnamefont{M.}~\bibnamefont{Haridas}},
  \bibinfo{author}{\bibfnamefont{J.}~\bibnamefont{Basu}},
  \bibinfo{author}{\bibfnamefont{A.}~\bibnamefont{Tiwari}}, \bibnamefont{and}
  \bibinfo{author}{\bibfnamefont{M.}~\bibnamefont{Venkatapathi}},
  \bibinfo{journal}{J. Appl. Phys.} \textbf{\bibinfo{volume}{114}},
  \bibinfo{pages}{064305} (\bibinfo{year}{2013}).

\bibitem[{\citenamefont{Haridas et~al.}(2011)\citenamefont{Haridas, Tripathi,
  and Basu}}]{haridas2011photoluminescence}
\bibinfo{author}{\bibfnamefont{M.}~\bibnamefont{Haridas}},
  \bibinfo{author}{\bibfnamefont{L.}~\bibnamefont{Tripathi}}, \bibnamefont{and}
  \bibinfo{author}{\bibfnamefont{J.}~\bibnamefont{Basu}},
  \bibinfo{journal}{Appl. Phys. Lett.} \textbf{\bibinfo{volume}{98}},
  \bibinfo{pages}{27} (\bibinfo{year}{2011}).

\bibitem[{\citenamefont{Haridas et~al.}(2010)\citenamefont{Haridas, Basu,
  Gosztola, and Wiederrecht}}]{haridas2010photoluminescence}
\bibinfo{author}{\bibfnamefont{M.}~\bibnamefont{Haridas}},
  \bibinfo{author}{\bibfnamefont{J.}~\bibnamefont{Basu}},
  \bibinfo{author}{\bibfnamefont{D.}~\bibnamefont{Gosztola}}, \bibnamefont{and}
  \bibinfo{author}{\bibfnamefont{G.}~\bibnamefont{Wiederrecht}},
  \bibinfo{journal}{Appl. Phys. Lett.} \textbf{\bibinfo{volume}{97}},
  \bibinfo{pages}{189} (\bibinfo{year}{2010}).

\bibitem[{\citenamefont{Haridas and Basu}(2010)}]{haridas2010controlled}
\bibinfo{author}{\bibfnamefont{M.}~\bibnamefont{Haridas}} \bibnamefont{and}
  \bibinfo{author}{\bibfnamefont{J.}~\bibnamefont{Basu}},
  \bibinfo{journal}{Nanotechnol.} \textbf{\bibinfo{volume}{21}},
  \bibinfo{pages}{415202} (\bibinfo{year}{2010}).

\bibitem[{\citenamefont{Kovacs et~al.}(1986)\citenamefont{Kovacs, Loutfy,
  Vincett, Jennings, and Aroca}}]{kovacs1986distance}
\bibinfo{author}{\bibfnamefont{G.}~\bibnamefont{Kovacs}},
  \bibinfo{author}{\bibfnamefont{R.}~\bibnamefont{Loutfy}},
  \bibinfo{author}{\bibfnamefont{P.}~\bibnamefont{Vincett}},
  \bibinfo{author}{\bibfnamefont{C.}~\bibnamefont{Jennings}}, \bibnamefont{and}
  \bibinfo{author}{\bibfnamefont{R.}~\bibnamefont{Aroca}},
  \bibinfo{journal}{Langmuir} \textbf{\bibinfo{volume}{2}},
  \bibinfo{pages}{689} (\bibinfo{year}{1986}).

\bibitem[{\citenamefont{Ye et~al.}(1997)\citenamefont{Ye, Fang, and
  Sun}}]{ye1997surface}
\bibinfo{author}{\bibfnamefont{Q.}~\bibnamefont{Ye}},
  \bibinfo{author}{\bibfnamefont{J.}~\bibnamefont{Fang}}, \bibnamefont{and}
  \bibinfo{author}{\bibfnamefont{L.}~\bibnamefont{Sun}}, \bibinfo{journal}{J.
  Phys. Chem. B} \textbf{\bibinfo{volume}{101}}, \bibinfo{pages}{8221}
  (\bibinfo{year}{1997}).

\bibitem[{\citenamefont{McFarland et~al.}(2005)\citenamefont{McFarland, Young,
  Dieringer, and Van~Duyne}}]{mcfarland2005wavelength}
\bibinfo{author}{\bibfnamefont{A.~D.} \bibnamefont{McFarland}},
  \bibinfo{author}{\bibfnamefont{M.~A.} \bibnamefont{Young}},
  \bibinfo{author}{\bibfnamefont{J.~A.} \bibnamefont{Dieringer}},
  \bibnamefont{and} \bibinfo{author}{\bibfnamefont{R.~P.}
  \bibnamefont{Van~Duyne}}, \bibinfo{journal}{J. Phys. Chem. B}
  \textbf{\bibinfo{volume}{109}}, \bibinfo{pages}{11279}
  (\bibinfo{year}{2005}).

\bibitem[{\citenamefont{Dieringer et~al.}(2006)\citenamefont{Dieringer,
  McFarland, Shah, Stuart, Whitney, Yonzon, Young, Zhang, and
  Van~Duyne}}]{dieringer2006introductory}
\bibinfo{author}{\bibfnamefont{J.~A.} \bibnamefont{Dieringer}},
  \bibinfo{author}{\bibfnamefont{A.~D.} \bibnamefont{McFarland}},
  \bibinfo{author}{\bibfnamefont{N.~C.} \bibnamefont{Shah}},
  \bibinfo{author}{\bibfnamefont{D.~A.} \bibnamefont{Stuart}},
  \bibinfo{author}{\bibfnamefont{A.~V.} \bibnamefont{Whitney}},
  \bibinfo{author}{\bibfnamefont{C.~R.} \bibnamefont{Yonzon}},
  \bibinfo{author}{\bibfnamefont{M.~A.} \bibnamefont{Young}},
  \bibinfo{author}{\bibfnamefont{X.}~\bibnamefont{Zhang}}, \bibnamefont{and}
  \bibinfo{author}{\bibfnamefont{R.~P.} \bibnamefont{Van~Duyne}},
  \bibinfo{journal}{Faraday Discussions} \textbf{\bibinfo{volume}{132}},
  \bibinfo{pages}{9} (\bibinfo{year}{2006}).

\bibitem[{\citenamefont{Wang et~al.}(2013)\citenamefont{Wang, Zhu, Best,
  Camden, and Crozier}}]{wang2013wafer}
\bibinfo{author}{\bibfnamefont{D.}~\bibnamefont{Wang}},
  \bibinfo{author}{\bibfnamefont{W.}~\bibnamefont{Zhu}},
  \bibinfo{author}{\bibfnamefont{M.~D.} \bibnamefont{Best}},
  \bibinfo{author}{\bibfnamefont{J.~P.} \bibnamefont{Camden}},
  \bibnamefont{and} \bibinfo{author}{\bibfnamefont{K.~B.}
  \bibnamefont{Crozier}}, \bibinfo{journal}{Sci. Rep.}
  \textbf{\bibinfo{volume}{3}}, \bibinfo{pages}{2867} (\bibinfo{year}{2013}).

\bibitem[{\citenamefont{Gabudean et~al.}(2012)\citenamefont{Gabudean, Focsan,
  and Astilean}}]{gabudean2012gold}
\bibinfo{author}{\bibfnamefont{A.~M.} \bibnamefont{Gabudean}},
  \bibinfo{author}{\bibfnamefont{M.}~\bibnamefont{Focsan}}, \bibnamefont{and}
  \bibinfo{author}{\bibfnamefont{S.}~\bibnamefont{Astilean}},
  \bibinfo{journal}{J. Phys. Chem. C} \textbf{\bibinfo{volume}{116}},
  \bibinfo{pages}{12240} (\bibinfo{year}{2012}).

\bibitem[{\citenamefont{Ivanov et~al.}(2016)\citenamefont{Ivanov, Todorov,
  Petrov, Ritacco, Giocondo, and Vlakhov}}]{ivanov2016strong}
\bibinfo{author}{\bibfnamefont{V.}~\bibnamefont{Ivanov}},
  \bibinfo{author}{\bibfnamefont{N.}~\bibnamefont{Todorov}},
  \bibinfo{author}{\bibfnamefont{L.}~\bibnamefont{Petrov}},
  \bibinfo{author}{\bibfnamefont{T.}~\bibnamefont{Ritacco}},
  \bibinfo{author}{\bibfnamefont{M.}~\bibnamefont{Giocondo}}, \bibnamefont{and}
  \bibinfo{author}{\bibfnamefont{E.}~\bibnamefont{Vlakhov}}, in
  \emph{\bibinfo{booktitle}{J. Phys.: Conference Series}}
  (\bibinfo{organization}{IOP Publishing}, \bibinfo{year}{2016}), vol.
  \bibinfo{volume}{764}, p. \bibinfo{pages}{012023}.

\bibitem[{\citenamefont{Kennedy et~al.}(1999)\citenamefont{Kennedy, Spaeth,
  Dickey, and Carron}}]{kennedy1999determination}
\bibinfo{author}{\bibfnamefont{B.}~\bibnamefont{Kennedy}},
  \bibinfo{author}{\bibfnamefont{S.}~\bibnamefont{Spaeth}},
  \bibinfo{author}{\bibfnamefont{M.}~\bibnamefont{Dickey}}, \bibnamefont{and}
  \bibinfo{author}{\bibfnamefont{K.}~\bibnamefont{Carron}},
  \bibinfo{journal}{J. Phys. Chem. B} \textbf{\bibinfo{volume}{103}},
  \bibinfo{pages}{3640} (\bibinfo{year}{1999}).


\bibitem[{\citenamefont{Le~Ru and Etchegoin}(2006)}]{le2006rigorous}
\bibinfo{author}{\bibfnamefont{E.}~\bibnamefont{Le~Ru}} \bibnamefont{and}
  \bibinfo{author}{\bibfnamefont{P.}~\bibnamefont{Etchegoin}},
  \bibinfo{journal}{Chem. Phys. Lett.} \textbf{\bibinfo{volume}{423}},
  \bibinfo{pages}{63} (\bibinfo{year}{2006}).
  
 \bibitem[{\citenamefont{fromm et~al.}(2006)}]{fromm2006SERS}
\bibinfo{author}{\bibfnamefont{D.}~\bibnamefont{Fromm}} \bibnamefont{and}
  \bibinfo{author}{\bibfnamefont{A.}~\bibnamefont{Kinkhabwala}},
  \bibinfo{author}{\bibfnamefont{P.}~\bibnamefont{Schuck}},
  \bibinfo{author}{\bibfnamefont{W.}~\bibnamefont{Moerner}},
  \bibinfo{author}{\bibfnamefont{A.}~\bibnamefont{Sundaramurthy}},
  bibnamefont{and}
  \bibinfo{author}{\bibfnamefont{G.}~\bibnamefont{Kino}},
  \bibinfo{journal}{J. Chem. Phys.} \textbf{\bibinfo{volume}{124}},
  \bibinfo{pages}{61101} (\bibinfo{year}{2006}).


\bibitem[{\citenamefont{Kneipp}(2016)}]{kneipp2016SERSrigorous}
\bibinfo{author}{\bibfnamefont{K.}~\bibnamefont{Kneipp}}
  \bibinfo{journal}{J. Phys. Chem. C} \textbf{\bibinfo{volume}{120}},
  \bibinfo{pages}{21076} (\bibinfo{year}{2016}).
  
  
 \bibitem[{\citenamefont{Sharma, Frontiera, Henry, Ringe, and Van Duyne}(2012)}]{sharma2012SERSreview}
\bibinfo{author}{\bibfnamefont{B.}~\bibnamefont{Sharma}} \bibnamefont{and}
  \bibinfo{author}{\bibfnamefont{R.}~\bibnamefont{Frontiera}},
  \bibinfo{author}{\bibfnamefont{A.}~\bibnamefont{Henry}},
  \bibinfo{author}{\bibfnamefont{E.}~\bibnamefont{Ringe}},\bibnamefont{and}
  \bibinfo{author}{\bibfnamefont{R.}~\bibnamefont{Van Duyne}},
  \bibinfo{journal}{Materials Today} \textbf{\bibinfo{volume}{15}},
  \bibinfo{pages}{16} (\bibinfo{year}{2012}).
  

\bibitem[{\citenamefont{Raimond et~al.}(1982)\citenamefont{Raimond, Goy, Gross,
  Fabre, and Haroche}}]{raimond1982statistics}
\bibinfo{author}{\bibfnamefont{J.}~\bibnamefont{Raimond}},
  \bibinfo{author}{\bibfnamefont{P.}~\bibnamefont{Goy}},
  \bibinfo{author}{\bibfnamefont{M.}~\bibnamefont{Gross}},
  \bibinfo{author}{\bibfnamefont{C.}~\bibnamefont{Fabre}}, \bibnamefont{and}
  \bibinfo{author}{\bibfnamefont{S.}~\bibnamefont{Haroche}},
  \bibinfo{journal}{Phys. Rev. Lett.} \textbf{\bibinfo{volume}{49}},
  \bibinfo{pages}{1924} (\bibinfo{year}{1982}).

\bibitem[{\citenamefont{Zhu et~al.}(2015)\citenamefont{Zhu, Gauthier, Morin, Wu, Carmichael, and Mossberg}}]{zhu1990vacuum}
\bibinfo{author}{\bibfnamefont{Y.}~\bibnamefont{Zhu}},
  \bibinfo{author}{\bibfnamefont{D.}~\bibnamefont{Gauthier}},
  \bibinfo{author}{\bibfnamefont{S.}~\bibnamefont{Morin}},
  \bibinfo{author}{\bibfnamefont{Q.}~\bibnamefont{Wu}},
  \bibinfo{author}{\bibfnamefont{H.}~\bibnamefont{Carmichael}},
  \bibnamefont{and} \bibinfo{author}{\bibfnamefont{T.}~\bibnamefont{Mossberg}},
  \bibinfo{journal}{Phys. Rev. Lett.} \textbf{\bibinfo{volume}{64}},
  \bibinfo{pages}{2499} (\bibinfo{year}{1990}).
  
\bibitem[{\citenamefont{Boca et~al.}(2004)\citenamefont{Boca, Miller, Birnbaum, Boozer, McKeever, and Kimble}}]{boca2004observation}
\bibinfo{author}{\bibfnamefont{A.}~\bibnamefont{Boca}},
  \bibinfo{author}{\bibfnamefont{R.}~\bibnamefont{Miller}},
  \bibinfo{author}{\bibfnamefont{K.}~\bibnamefont{Birnbaum}},
  \bibinfo{author}{\bibfnamefont{A.}~\bibnamefont{Boozer}},
  \bibinfo{author}{\bibfnamefont{J.}~\bibnamefont{McKeever}},
  \bibnamefont{and} \bibinfo{author}{\bibfnamefont{H.}~\bibnamefont{Kimble}},
  \bibinfo{journal}{Phys. Rev. Lett.} \textbf{\bibinfo{volume}{93}},
  \bibinfo{pages}{233603} (\bibinfo{year}{2004}).
  
  \bibitem[{\citenamefont{Yoshie et~al.}(2004)\citenamefont{Yoshie, Scherer, Hendrickson, Khitrova, Gibbs, Rupper, Ell, Shchekin, and Deppe}}]{yoshie2004vacuum}
\bibinfo{author}{\bibfnamefont{T.}~\bibnamefont{Yoshie}},
  \bibinfo{author}{\bibfnamefont{A.}~\bibnamefont{Scherer}},
  \bibinfo{author}{\bibfnamefont{J.}~\bibnamefont{Hendrickson}},
  \bibinfo{author}{\bibfnamefont{G.}~\bibnamefont{Khitrova}},
  \bibinfo{author}{\bibfnamefont{H.}~\bibnamefont{Gibbs}},
  \bibinfo{author}{\bibfnamefont{G.}~\bibnamefont{Rupper}},
  \bibinfo{author}{\bibfnamefont{C.}~\bibnamefont{Ell}},
  \bibinfo{author}{\bibfnamefont{O.}~\bibnamefont{Shchekin}},
  \bibnamefont{and} \bibinfo{author}{\bibfnamefont{D.}~\bibnamefont{Deppe}},
  \bibinfo{journal}{Nature} \textbf{\bibinfo{volume}{432}},
  \bibinfo{pages}{200} (\bibinfo{year}{2004}).

\bibitem[{\citenamefont{Bernardot et~al.}(1992)\citenamefont{Bernardot, Nussenzveig, Brune, Raimond, and Haroche}}]{bernardot1992vacuum}
\bibinfo{author}{\bibfnamefont{F.}~\bibnamefont{Bernardot}},
  \bibinfo{author}{\bibfnamefont{P.}~\bibnamefont{Nussenzveig}},
  \bibinfo{author}{\bibfnamefont{M.}~\bibnamefont{Brune}},
  \bibinfo{author}{\bibfnamefont{J.}~\bibnamefont{Raimond}},
  \bibnamefont{and} \bibinfo{author}{\bibfnamefont{S.}~\bibnamefont{Haroche}},
  \bibinfo{journal}{Eur. Phys. Lett.} \textbf{\bibinfo{volume}{17}},
  \bibinfo{pages}{33} (\bibinfo{year}{1992}).

\bibitem[{\citenamefont{Bellessa, Bonnand, Plenet, and Mugnier}(2004)\citenamefont{Bellessa, Bonnand, Plenet, and Mugnier}}]{bellessa2004matter}
\bibinfo{author}{\bibfnamefont{J.}~\bibnamefont{Bellessa}},
  \bibinfo{author}{\bibfnamefont{C.}~\bibnamefont{Bonnand}},
  \bibinfo{author}{\bibfnamefont{J.}~\bibnamefont{Plenet}},
  \bibnamefont{and} \bibinfo{author}{\bibfnamefont{J.}~\bibnamefont{Mugnier}},
  \bibinfo{journal}{Phys. Rev. Lett.} \textbf{\bibinfo{volume}{93}},
  \bibinfo{pages}{036404} (\bibinfo{year}{2004}).
  
\bibitem[{\citenamefont{Zengin et~al.}(2015)\citenamefont{Zengin, Wersäll, Nilsson, Antosiewicz, Käll, and Shegai}}]{zengin2015matter}
\bibinfo{author}{\bibfnamefont{G.}~\bibnamefont{Zengin}},
  \bibinfo{author}{\bibfnamefont{M.}~\bibnamefont{Wersäll}},
  \bibinfo{author}{\bibfnamefont{S.}~\bibnamefont{Nilsson}},
  \bibinfo{author}{\bibfnamefont{T.}~\bibnamefont{Antosiewicz}},
  \bibinfo{author}{\bibfnamefont{M.}~\bibnamefont{Käll}},
  \bibnamefont{and} \bibinfo{author}{\bibfnamefont{T.}~\bibnamefont{Shegai}},
  \bibinfo{journal}{Phys. Rev. Lett.} \textbf{\bibinfo{volume}{114}},
  \bibinfo{pages}{157401} (\bibinfo{year}{2015}).
  

\bibitem[{\citenamefont{Adam, Salomon, de Fornel, and Goudonnet}(1993)\citenamefont{Adam, Salomon, de Fornel, and Goudonnet}}]{adam93evanescent}
\bibinfo{author}{\bibfnamefont{P.}~\bibnamefont{Adam}},
  \bibinfo{author}{\bibfnamefont{L.}~\bibnamefont{Salomon}},
  \bibinfo{author}{\bibfnamefont{F.}~\bibnamefont{de Fornel}},
  \bibnamefont{and} \bibinfo{author}{\bibfnamefont{J.}~\bibnamefont{Goudonnet}},
  \bibinfo{journal}{Phys. Rev. B} \textbf{\bibinfo{volume}{48}},
  \bibinfo{pages}{2680} (\bibinfo{year}{1993}).
  
\bibitem[{\citenamefont{Meixner, Bopp, and Tarrach}(1994)\citenamefont{Meixner, Bopp, and Tarrach}}]{meixner94evanescent}
\bibinfo{author}{\bibfnamefont{A.}~\bibnamefont{Meixner}},
  \bibinfo{author}{\bibfnamefont{M.}~\bibnamefont{Bopp}},
  \bibnamefont{and} \bibinfo{author}{\bibfnamefont{G.}~\bibnamefont{Tarrach}},
  \bibinfo{journal}{Appl. Opt.} \textbf{\bibinfo{volume}{33}},
  \bibinfo{pages}{7995} (\bibinfo{year}{1994}).
 
\bibitem[{\citenamefont{Baena, Jelinek, Marqués, and Medina}(2005)\citenamefont{Baena, Jelinek, Marqués, and Medina}}]{baena05evanescent}
\bibinfo{author}{\bibfnamefont{J.}~\bibnamefont{Baena}},
  \bibinfo{author}{\bibfnamefont{L.}~\bibnamefont{Jelinek}},
  \bibinfo{author}{\bibfnamefont{R.}~\bibnamefont{Marqués}},
  \bibnamefont{and} \bibinfo{author}{\bibfnamefont{F.}~\bibnamefont{Medina}},
  \bibinfo{journal}{Phys. Rev. B} \textbf{\bibinfo{volume}{72}},
  \bibinfo{pages}{075116} (\bibinfo{year}{2005}).
 
\bibitem[{\citenamefont{Loomis and Maris}(1994)\citenamefont{Loomis and Maris}}]{loomis94fluctuation}
\bibinfo{author}{\bibfnamefont{J.}~\bibnamefont{Loomis}},
  \bibnamefont{and} \bibinfo{author}{\bibfnamefont{H.}~\bibnamefont{Maris}},
  \bibinfo{journal}{Phys. Rev. B} \textbf{\bibinfo{volume}{50}},
  \bibinfo{pages}{18517} (\bibinfo{year}{1994}). 

\bibitem[{\citenamefont{Rytov}(1953)\citenamefont{Rytov}}]{rytov53fluctuation}
\bibinfo{author}{\bibfnamefont{S.}~\bibnamefont{Rytov}},
  \bibinfo{journal}{ Theory of Electrical Fluctuations and Thermal Radiation, Academy of Sciences Press, Moscow, 1953.} 

\bibitem[{\citenamefont{Parikh and Wilczek}(2000)\citenamefont{Parikh and Wilczek}}]{Hawking_radiation}
\bibinfo{author}{\bibfnamefont{M.}~\bibnamefont{Parikh}},
  \bibinfo{author}{\bibfnamefont{F.}~\bibnamefont{Wilczek}},
  \bibinfo{journal}{Phys. Rev. Lett.} \textbf{\bibinfo{volume}{85}},
  \bibinfo{pages}{5042} (\bibinfo{year}{2000}).


\bibitem[{Sup(2018)}]{Supplementary}
\bibinfo{journal}{Supplementary material at [URL of Journal] for additional information on used methods and cited experimental results}
  (\bibinfo{year}{2018}).

\bibitem[{\citenamefont{Pustovit and
  Shahbazyan}(2009)}]{pustovit2009cooperative}
\bibinfo{author}{\bibfnamefont{V.~N.} \bibnamefont{Pustovit}} \bibnamefont{and}
  \bibinfo{author}{\bibfnamefont{T.~V.} \bibnamefont{Shahbazyan}},
  \bibinfo{journal}{Phys. Rev. Lett.} \textbf{\bibinfo{volume}{102}},
  \bibinfo{pages}{077401} (\bibinfo{year}{2009}).

\bibitem[{\citenamefont{Potma and Wiersma}(1998)}]{potma1998exciton}
\bibinfo{author}{\bibfnamefont{E.~O.} \bibnamefont{Potma}} \bibnamefont{and}
  \bibinfo{author}{\bibfnamefont{D.~A.} \bibnamefont{Wiersma}},
  \bibinfo{journal}{J. Chem. Phys.} \textbf{\bibinfo{volume}{108}},
  \bibinfo{pages}{4894} (\bibinfo{year}{1998}).

\bibitem[{\citenamefont{Gross and Haroche}(1982)}]{gross1982superradiance}
\bibinfo{author}{\bibfnamefont{M.}~\bibnamefont{Gross}} \bibnamefont{and}
  \bibinfo{author}{\bibfnamefont{S.}~\bibnamefont{Haroche}},
  \bibinfo{journal}{Phys. Rep.} \textbf{\bibinfo{volume}{93}},
  \bibinfo{pages}{301} (\bibinfo{year}{1982}).

\bibitem[{\citenamefont{Scheibner et~al.}(2007)\citenamefont{Scheibner,
  Schmidt, Worschech, Forchel, Bacher, Passow, and
  Hommel}}]{scheibner2007superradiance}
\bibinfo{author}{\bibfnamefont{M.}~\bibnamefont{Scheibner}},
  \bibinfo{author}{\bibfnamefont{T.}~\bibnamefont{Schmidt}},
  \bibinfo{author}{\bibfnamefont{L.}~\bibnamefont{Worschech}},
  \bibinfo{author}{\bibfnamefont{A.}~\bibnamefont{Forchel}},
  \bibinfo{author}{\bibfnamefont{G.}~\bibnamefont{Bacher}},
  \bibinfo{author}{\bibfnamefont{T.}~\bibnamefont{Passow}}, \bibnamefont{and}
  \bibinfo{author}{\bibfnamefont{D.}~\bibnamefont{Hommel}},
  \bibinfo{journal}{Nat. Phys.} \textbf{\bibinfo{volume}{3}},
  \bibinfo{pages}{106} (\bibinfo{year}{2007}).

\bibitem[{\citenamefont{Goban et~al.}(2015)\citenamefont{Goban, Hung, Hood, Yu,
  Muniz, Painter, and Kimble}}]{Goban2015superradiance}
\bibinfo{author}{\bibfnamefont{A.}~\bibnamefont{Goban}},
  \bibinfo{author}{\bibfnamefont{C.~L.} \bibnamefont{Hung}},
  \bibinfo{author}{\bibfnamefont{J.~D.} \bibnamefont{Hood}},
  \bibinfo{author}{\bibfnamefont{S.~P.} \bibnamefont{Yu}},
  \bibinfo{author}{\bibfnamefont{J.~A.} \bibnamefont{Muniz}},
  \bibinfo{author}{\bibfnamefont{O.}~\bibnamefont{Painter}}, \bibnamefont{and}
  \bibinfo{author}{\bibfnamefont{H.~J.} \bibnamefont{Kimble}},
  \bibinfo{journal}{Phys. Rev. Lett.} \textbf{\bibinfo{volume}{115}},
  \bibinfo{pages}{063601} (\bibinfo{year}{2015}).

\bibitem[{\citenamefont{Juan et~al.}(2017)\citenamefont{Juan, Bradac, Besga,
  Molina-Terriza, and Volz}}]{Juan2017cooperative}
\bibinfo{author}{\bibfnamefont{M.~L.} \bibnamefont{Juan}},
  \bibinfo{author}{\bibfnamefont{C.}~\bibnamefont{Bradac}},
  \bibinfo{author}{\bibfnamefont{B.}~\bibnamefont{Besga}},
  \bibinfo{author}{\bibfnamefont{G.}~\bibnamefont{Molina-Terriza}},
  \bibnamefont{and} \bibinfo{author}{\bibfnamefont{T.}~\bibnamefont{Volz}},
  \bibinfo{journal}{Nat. Phys.} \textbf{\bibinfo{volume}{13}},
  \bibinfo{pages}{241} (\bibinfo{year}{2017}).

\bibitem[{\citenamefont{Praveena et~al.}(2015)\citenamefont{Praveena,
  Mukherjee, Venkatapathi, and Basu}}]{praveena2015plasmon}
\bibinfo{author}{\bibfnamefont{M.}~\bibnamefont{Praveena}},
  \bibinfo{author}{\bibfnamefont{A.}~\bibnamefont{Mukherjee}},
  \bibinfo{author}{\bibfnamefont{M.}~\bibnamefont{Venkatapathi}},
  \bibnamefont{and} \bibinfo{author}{\bibfnamefont{J.}~\bibnamefont{Basu}},
  \bibinfo{journal}{Phys. Rev. B} \textbf{\bibinfo{volume}{92}},
  \bibinfo{pages}{235403} (\bibinfo{year}{2015}).


\bibitem[{\citenamefont{Pustovit and Shahbazyan}(2010)}]{pustovit2010plasmon}
\bibinfo{author}{\bibfnamefont{V.~N.} \bibnamefont{Pustovit}} \bibnamefont{and}
  \bibinfo{author}{\bibfnamefont{T.~V.} \bibnamefont{Shahbazyan}},
  \bibinfo{journal}{Phys. Rev. B} \textbf{\bibinfo{volume}{82}},
  \bibinfo{pages}{075429} (\bibinfo{year}{2010}).

\bibitem[{\citenamefont{Van~Vlack et~al.}(2012)\citenamefont{Van~Vlack,
  Kristensen, and Hughes}}]{van2012spontaneous}
\bibinfo{author}{\bibfnamefont{C.}~\bibnamefont{Van~Vlack}},
  \bibinfo{author}{\bibfnamefont{P.~T.} \bibnamefont{Kristensen}},
  \bibnamefont{and} \bibinfo{author}{\bibfnamefont{S.}~\bibnamefont{Hughes}},
  \bibinfo{journal}{Phys. Rev. B} \textbf{\bibinfo{volume}{85}},
  \bibinfo{pages}{075303} (\bibinfo{year}{2012}).

\bibitem[{\citenamefont{Venkatapathi}(2014)}]{venkatapathi2014collective}
\bibinfo{author}{\bibfnamefont{M.}~\bibnamefont{Venkatapathi}},
  \bibinfo{journal}{J. Opt. Soc. Am. B} \textbf{\bibinfo{volume}{31}},
  \bibinfo{pages}{3153} (\bibinfo{year}{2014}).


\end{thebibliography}

\end{document}